\documentclass[twocolumn,letterpaper,aps,prl,superscriptaddress,showpacs,floatfix]{revtex4-1}

\usepackage{amsmath}
\usepackage{graphicx}	

\usepackage{xspace}	

\begin{document}

%
\title{ Finite-Size Scaling  of  Non-Gaussian Fluctuations Near the QCD Critical Point }

%
%
%
\author{ Roy~A.~Lacey}
\email[E-mail: ]{Roy.Lacey@Stonybrook.edu}
\affiliation{Department of Chemistry, 
Stony Brook University, 
Stony Brook, NY, 11794-3400, USA}
\affiliation{Dept. of Physics, 
Stony Brook University, 
Stony Brook, NY, 11794, USA}

\author{Peifeng Liu} 
\affiliation{Department of Chemistry, 
Stony Brook University, 
Stony Brook, NY, 11794-3400, USA}
\affiliation{Dept. of Physics, 
Stony Brook University, 
Stony Brook, NY, 11794, USA}
\author{ Niseem Magdy} 
\affiliation{Department of Chemistry, 
Stony Brook University, 
Stony Brook, NY, 11794-3400, USA}
\author{B. Schweid} 
\affiliation{Department of Chemistry, 
Stony Brook University, 
Stony Brook, NY, 11794-3400, USA}
\affiliation{Dept. of Physics, 
Stony Brook University, 
Stony Brook, NY, 11794, USA}
\author{ N.~N.~Ajitanand} 
\affiliation{Department of Chemistry, 
Stony Brook University, 
Stony Brook, NY, 11794-3400, USA}

\date{\today}

\begin{abstract}
An effective Finite-Size Scaling (FSS) of moment products from recent STAR measurements of  the variance $\sigma$, 
skewness $S$ and kurtosis $\kappa$ of net-proton multiplicity distributions, are reported  for a broad range of collision 
centralities in Au+Au ($\sqrt{s_{NN}}= 7.7 - 200$ GeV)  collisions. The products $S\sigma $  and  $\kappa \sigma^2 $, 
which are directly related to the hgher-order baryon number susceptibility ratios $\chi^{(3)}_B/\chi^{(2)}_B$ 
and $\chi^{(4)}_B/\chi^{(2)}_B$, show scaling patterns consistent with earlier indications for a second order phase transition 
at a critical end point (CEP)  in the plane of temperature  vs. baryon chemical potential ($T,\mu_B$) of the QCD phase diagram. 
The resulting scaling functions validate the earlier estimates of $T^{\text{cep}} \sim 165$~MeV and $\mu_B^{\text{cep}} \sim 95$~MeV 
for the location of the CEP,  and  the critical exponents  used to assign its 3D Ising model universality class.
\end{abstract}

\pacs{25.75.Dw} 
	



\maketitle
 

%
A major experimental theme at both the Super Proton Synchrotron (SPS)~\cite{Anticic:2009pe}   
and  the Relativistic Heavy Ion Collider (RHIC)~\cite{Caines:2009yu},
is the the study of observables  that could signal the location and character of the 
critical end point (CEP) -- the end point of  the first-order coexistence curve in the temperature vs. baryon chemical potential ($T, \mu_B$)
plane of the phase diagram for Quantum Chromodynamics (QCD) \cite{Itoh:1970,Shuryak:1983zb,Asakawa:1989bq,Stephanov:1998dy}.
The first beam energy scan at RHIC ($\sqrt{s_{NN}}= 7.7 - 200$ GeV -- BES-I) facilitated  measurements spanning a broad 
domain ($20< \mu_B <420$ MeV \cite{Cleymans:2006qe,Andronic:2009qf,Becattini:2012sq}) of this phase diagram.

A common strategy for locating the CEP is to scan the phase diagram by varying the beam collision energy ($\sqrt{s_{NN}}$), 
and look for non-monotonic behavior  of  experimental observables sensitive to the proximity of  reaction trajectories to the 
CEP~ \cite{Stephanov:1998dy,Stephanov:2004wx,Asakawa:2009aj,Stephanov:2008qz,Athanasiou:2010kw}.
This strategy stems from the fact that, for an infinite volume system, the critical point is characterized by several (power law) divergences 
linked to the divergence of  the correlation length $\xi \propto \left| t \right|^{-\nu} \equiv \left| T - T^{\mathrm{cep}} \right|^{-\nu}$. 
Notable examples are the quadratic variances of  event-by-event multiplicity distributions for net charge, baryon number, etc,
$\left< (\delta n) \right> \sim \xi^{\gamma/\nu}$, the isobaric heat capacity $C_p \sim \xi^{\gamma/\nu}$ and 
the isothermal compressibility $\kappa_T \sim \xi^{\gamma/\nu}$; the values of  the critical exponents 
$\nu$ and $\gamma$~\cite{exponents} can be determined via the static universality class of the CEP.

The higher-order moments of  the net-baryon multiplicity distributions, 
which are related to the higher-order baryon number susceptibilities, are predicted to be even more sensitive 
to the CEP since they are proportional to higher powers of  the correlation length \cite{Stephanov:2008qz,Athanasiou:2010kw}. 
Thus, the search for an increase/divergence or a non-monotonic trend in the excitation function 
of net-baryon (net-proton) fluctuations is a favored approach in ongoing experimental efforts to pinpoint the location 
of the CEP~\cite{Anticic:2009pe,Caines:2009yu,Adamczyk:2013dal,Luo:2014tga,Luo:2015doi}.

The observation of non-monotonic signatures, while important,  is neither necessary nor sufficient 
for identification and full characterization of the CEP.
Moreover, finite-size and finite-time effects impose non-negligible constraints on the magnitude of $\xi$ \cite{Berdnikov:1999ph}. 
If the effective correlation length associated with finite-time effects $\hat{\xi} \ge L$, a  pseudo-critical  point with 
$\xi  > L$, could result from the finite volume ($V \propto L^3$ ) systems created in heavy ion collisions. 
Such a pseudocritical point would lead to a characteristic power law volume dependence of the magnitude ($\chi ^{\text{max}}_{T}$), 
width ($\delta T$) and peak position ($\bar{\tau}_T$) of the susceptibility and its related observables \cite{Ladrem:2004dw};
\begin{eqnarray}
\chi ^{\text{max}}_T(V)  \sim L^{\gamma /\nu},          
\label{eq:1} \\
\delta T (V)  \sim  L^{- \frac{1}{\nu}},                                                                               
\label{eq:2} \\
\bar{\tau}_T(V) \sim  T^{\text{cep}}(V) - T^{\text{cep}}(\infty)   \sim  L^{- \frac{1}{\nu}},  
\label{eq:3}
\end{eqnarray}
where $\nu$ and $\gamma$ are critical exponents which characterize the divergence of 
$\xi$ and $\chi_T$ respectively.  Therefore, the reduction of the magnitude of  $\chi ^{\text{max}}_T(V)$
($\chi ^{\text{max}}_{\mu_B}(V)$ ), broadening of the transition region $\delta T (V)$ ($\delta \mu_B (V)$)
and the shift of  $T^{\text{cep}}$ ($\mu_B^{\text{cep}}$), which  all increase as the volume decreases, 
provide a unique set of finite-size signatures for locating and characterizing 
the CEP~\cite{Ladrem:2004dw,Palhares:2009tf,Lizhu:2010wy,Lacey:2014wqa,Lacey:2015yxg}.

In general,  the {\em i}th order susceptibility can be written in a  finite-time-finite-size scaling form as  \cite{Suzuki1977,XSChen1996,Kibble-Zurek};
\begin{eqnarray}
\chi^{(i )}(\tau, t, L)  \sim  L^3b^{i\delta\beta/\nu}\chi^{(i)}(\tau b^{1/\nu}, tb^{-z}, L^{-1}b),          
\label{eq:4}
\end{eqnarray}
where $b$ is the renormalization-group scale factor, $z$ is the dynamic critical exponent and  $\delta$ and $\beta$ are additional 
static critical exponents. A suitable choice of the scale factor $b$ gives 
\begin{eqnarray}
\chi^{(i )}  \sim  L^{3+i\delta\beta/\nu} f_{1}(\tau L^{1/\nu}, RL^{r}),          
\label{eq:5}
\end{eqnarray}
and 
\begin{eqnarray}
\chi^{(i )}  =  L^3R^{-i\delta\beta/r\nu} f_{2}(\tau R^{-1/r\nu}, L^{-1}R^{-1/r}),          
\label{eq:6}
\end{eqnarray}
for the Finite-Size Scaling (FSS) and  Finite-Time Scaling (FTS) forms respectively. 
Here, $r = z + 1/\nu$ and $R$  can be associated with a cooling rate which drives the system from an initial 
temperature $T_i$ through the critical temperature $T_{\text{cep}}$  \cite{Kibble-Zurek}.

The regime of FSS is satisfied for $L^{-1}R^{-1/r} >> 1$ and $\tau R^{-1/\nu r}$ = $\tau L^{1/\nu}(L^{-1}R^{-1/r})^{1/\nu} << 1$,
and the ratio of susceptibilities can be expressed as;
\begin{eqnarray}
\frac{\chi^{(n)}}{\chi^{(m)}}  =  L^{(n-m)\delta\beta/\nu} f(\tau L^{1/\nu}).          
\label{eq:7}
\end{eqnarray}
The regime of FTS is restricted to $L^{-1}R^{-1/r} << 1$ or  $R^{-1/r} << L$. 
Therefore, if finite-time effects dominate, $\xi > L > \hat{\xi}$ and  ${\chi^{(n)}}/{\chi^{(m)}}$ will be 
essentially independent of  $L$ ({\em i.e.} ${(n-m)\delta\beta/\nu} \sim 0$) and $\chi^{(i )}  \sim  L^3$.
Eq.\ref{eq:7} implies that an observable related to the susceptibility ratio ${\chi^{(n)}}/{\chi^{(m)}}$,  
obtained for different system sizes, can be re-scaled to the non-singular scaling function $f(\tau L^{1/\nu})$. 
That is, FSS for different values of $L$, should lead to data collapse onto a single curve 
for robust values of $T^{\text{cep}}$,  $\mu_B^{\text{cep}}$ and the associated critical exponents;
\begin{equation}
\label{eq:8}
\begin{split}
\frac{\chi^{(n)}}{\chi^{(m)}}L^{(m-n)\delta\beta/\nu}  \text{  vs.  }  t_T L^{1/\nu},  
\\
\frac{\chi^{(n)}}{\chi^{(m)}}L^{(m-n)\delta\beta/\nu}  \text{  vs.  }  t_{\mu_B} L^{1/\nu},  
\end{split}
\end{equation}
where $\tau_T = (T - T^{\text{cep}})/T^{\text{cep}}$ and  $\tau_{\mu_B} = (\mu_B - \mu_B^{\text{cep}})/\mu_B^{\text{cep}}$
are the reduced temperature and baryon chemical potential respectively. 
A further simplification of these expressions results from the influence of finite-time effects which render the exponent for $L$ on the left-hand side 
of  Eq.~\ref{eq:8}, ${(m-n)\delta\beta/\nu} \sim 0$.
%
%
\begin{figure*}[t]
\includegraphics[width=0.70\linewidth]{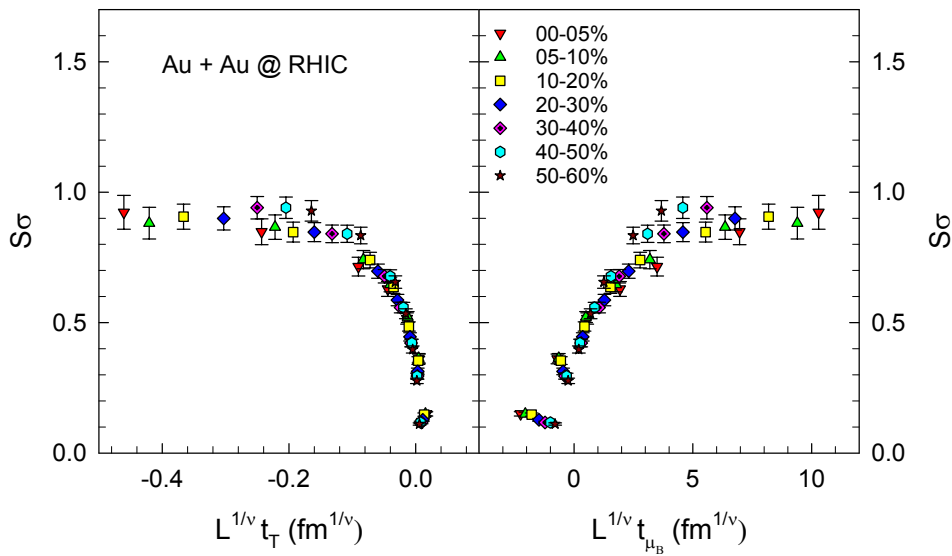}
\caption{$S\sigma$ vs. $L^{1/\nu} t_T$ (left panel)  and   $L^{1/\nu} t_{\mu_B}$ (right panel)
for 0-5\%,  5-10\%, 10-20\%, 20-30\%, 30-40\%, 40-50\% and 50-60\% Au+Au collisions
for $0.4 < p_T < 0.8$~GeV and $|y| < 0.5$. The efficiency corrected data are taken from 
Ref.~\cite{Adamczyk:2013dal}; scaling was performed with $T^{\text{cep}} \sim 165$~MeV, 
$\mu_B^{\text{cep}} \sim 95$~MeV and $\nu \sim 0.66$~\cite{Lacey:2014wqa,Lacey:2015yxg}.
}
\label{Fig1}
\vspace{10pt}
\end{figure*}

In recent work~\cite{Lacey:2014wqa,Lacey:2015yxg},  the centrality dependent excitation functions for the Gaussian emission 
source radii difference ($R^2_{\text{out}} - R^2_{\text{side}} = \Delta R^2$), obtained from  two-pion interferometry 
measurements in Au+Au ($\sqrt{s_{NN}}= 7.7 - 200$ GeV)  and Pb+Pb ($\sqrt{s_{NN}}= 2.76$ TeV) 
collisions, were employed to search for the CEP;  $\Delta R^2$ can be linked  to a susceptibility, given its 
connection to the compressibility~\cite{Lacey:2014wqa,Lacey:2015yxg}. 
The observed  non-monotonic excitation functions validated the 
characteristic finite-size scaling patterns expected for a deconfinement phase transition and the CEP.
The ensuing Finite-Size Scaling analysis of these data~\cite{Lacey:2014wqa,Lacey:2015yxg} indicated a second order 
phase transition with $T^{\text{cep}} \sim 165$~MeV and $\mu_B^{\text{cep}} \sim 95$~MeV 
 for the location of the critical end point, as well as critical exponents ($\nu = 0.66 \pm 0.05$ and $\gamma = 1.15 \pm 0.065$)
which placed the CEP in the 3D Ising model universality class.  However, it remains an open question as to whether 
these findings are consistent with the fluctuation measurements (especially net-proton fluctuations) obtained in the 
same experiments.

In this work, we investigate whether the estimates for $T^{\text{cep}}$,  $\mu_B^{\text{cep}}$ 
and the associated critical exponents reported in Refs.~\cite{Lacey:2014wqa,Lacey:2015yxg},  lead to robust Finite-Size Scaling 
of recent STAR measurements of  the beam energy and centrality dependent moment products $S\sigma $  and  $\kappa \sigma^2 $, 
for net-protons. Here,  $\sigma$, $S$ and  $\kappa$ are the variance, skewness and  kurtosis (respectively) of  the event-by-event 
net-proton multiplicity distributions. We find clear evidence  for an effective Finite-Size Scaling of  $S\sigma $  and  $\kappa \sigma^2 $,
compatible with the location and properties of the CEP  reported in Refs.~\cite{Lacey:2014wqa,Lacey:2015yxg}.
 
The efficiency corrected data employed in the present analysis, are obtained from centrality dependent 
fluctuation measurements ($0.4 < p_T < 0.8$~GeV/c and  $|y| < 0.5$) reported by the STAR 
collaboration for Au+Au collisions spanning the range $\sqrt{s_{NN}}= 7.7 - 200$ GeV~\cite{Adamczyk:2013dal}.
The moment products ($S\sigma $  and  $\kappa \sigma^2 $) which are obtained from  the cumulants of  the 
net-proton multiplicity distributions, $C_3/C_2 = S\sigma$ and $C_4/C_2 = \kappa\sigma^2$, 
are related to the ratio of  baryon number susceptibilities, $S\sigma \sim \chi^{(3)}_B/\chi^{(2)}_B$ 
and $\kappa\sigma^2  \sim \chi^{(4)}_B/\chi^{(2)}_B$  \cite{Ejiri:2005wq,Cheng:2008zh,Gupta:2011wh,Gavai:2010zn}. 
The systematic uncertainties for these measurements are  reported to be of order
4\% for  $\sigma$, 5\% for $S$ and 12\% for $\kappa$~\cite{Adamczyk:2013dal}.
The reported statistical uncertainties are in general, much larger for $\kappa\sigma^2$ than 
for $S\sigma$. Note as well that it is assumed that the shape of the net-proton distributions reflects the net-baryon
distributions up to distortions smaller than the estimated uncertainties for the cumulant measurements.

The size parameter ${\bar{R}} = L$, employed in our Finite-Size Scaling analysis 
is obtained from a Monte Carlo Glauber (MC-Glauber) calculation \cite{glauber,Lacey:2010hw,Adare:2013nff},
performed for several collision centralities at each beam energy. In each of these calculations, a subset of the nucleons 
become participants ($N_{\text{part}}$) in each collision by undergoing an initial inelastic N+N interaction.  
The transverse distribution of these participants in the X-Y  plane has RMS widths $\sigma_x$ and $\sigma_y$ 
along its principal axes. We define and compute $\bar{R}$, the characteristic initial transverse size, as 
${1}/{\bar{R}}~=~\sqrt{\left({1}/{\sigma_x^2}+{1}/{\sigma_y^2}\right)}$ \cite{Bhalerao:2005mm}.
Here, it  is noteworthy that the  three HBT radii  ${R_{\text{out}}}$,  $R_{\text{side}}$  and $R_{\text{long}}$,
which serve to characterize the space-time dimensions of the emitting sources in Au+Au collisions over the 
same span of beam energies, all show a linear dependence on  $\bar{R}$~\cite{Lacey:2014rxa,Adare:2014qvs}.
The systematic uncertainties for $\bar{R}$, obtained via variation of the model parameters, are 
less than 10\% \cite{Lacey:2010hw,Adare:2013nff}. 

%
%
\begin{figure*}[t]
\includegraphics[width=0.70\linewidth]{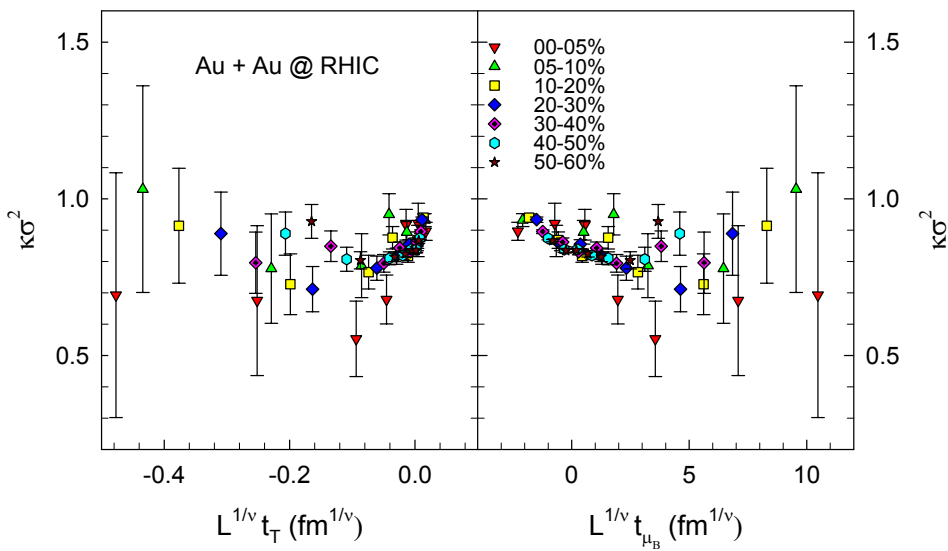}
\caption {$\kappa\sigma^2$ vs. $L^{1/\nu} t_T$ (left panel)  and   $L^{1/\nu} t_{\mu_B}$ (right panel)
for 0-5\%,  5-10\%, 10-20\%, 20-30\%, 30-40\%, 40-50\% and 50-60\% Au+Au collisions
for $0.4 < p_T < 0.8$~GeV and $|y| < 0.5$. The efficiency corrected data are taken from 
Ref.~\cite{Adamczyk:2013dal}; scaling was performed with $T^{\text{cep}} \sim 165$~MeV, 
$\mu_B^{\text{cep}} \sim 95$~MeV and $\nu \sim 0.66$~\cite{Lacey:2014wqa,Lacey:2015yxg}.
}
\label{Fig2}
\end{figure*}
%
%

%
%
\begin{figure*}[t]
\includegraphics[width=0.70\linewidth]{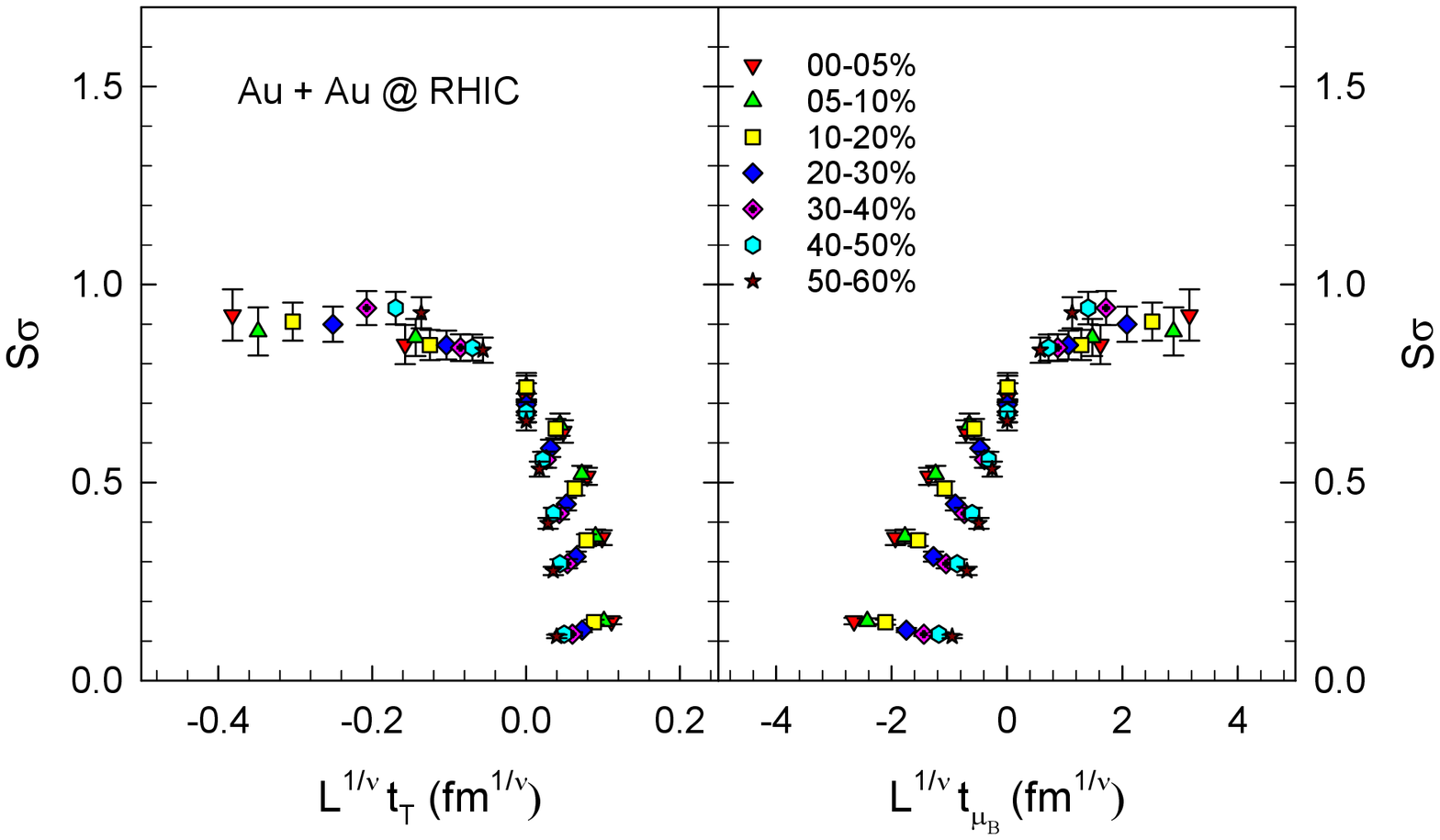}
\caption { Same as Fig.~\ref{Fig1},  for the test values $\acute{T}^{\text{cep}} \sim 160$~MeV and $\acute{\mu_B}^{\text{cep}} \sim 205$~MeV
($\sqrt{s_{NN}} \sim 20$ GeV).
}
\label{Fig3}
\end{figure*}

Validation tests for  finite-size scaling of  the moment products $S\sigma \sim \chi^{(3)}_B/\chi^{(2)}_B$ 
and $\kappa\sigma^2  \sim \chi^{(4)}_B/\chi^{(2)}_B$ were carried out for the full set of centrality dependent 
measurements as follows. First, we exploit the phenomenology of  thermal 
models \cite{Cleymans:2006qe,Andronic:2009qf,Becattini:2012sq} for the freeze-out
region and use the $\sqrt{s_{NN}}$ values to obtain the corresponding  $T$ and $\mu_B$ values \cite{Cleymans:2006qe}.
We then use the previously extracted~\cite{Lacey:2014wqa} values of $T^{\text{cep}} \sim 165$~MeV and $\mu_B^{\text{cep}} \sim 95$~MeV
to obtain $t_T$  and $t_{\mu_B}$ over the full range of the measurements.  
Subsequently, the value $\nu \sim 0.66$~\cite{Lacey:2014wqa} and  those for $\bar{R}$ ($L$) were used in conjunction  
with Eq.~\ref{eq:8} and ${(m-n)\delta\beta/\nu} \sim 0$, to test for collapse of the centrality dependent data onto a single curve. 
Here, ${(m-n)\delta\beta/\nu} \sim 0$ takes account of the strong influence of  finite-time effects, given the large value of $z = 3$, for the thermal 
slow mode which drives the critical dynamics associated with fluctuations \cite{Berdnikov:1999ph,Minami:2011un}. This influence was 
further validated via the scaling functions expected for $L^{-3}\chi^{(i )}$  vs.  $f(\tau L^{1/\nu})$.

Figures~\ref{Fig1} and \ref{Fig2} show the results of  the FSS tests for $S\sigma$ and $\kappa\sigma^2$ respectively. 
The left and right panel in each figure, show results for the scaled variable $t_TL^{1/\nu}$ and   $t_{\mu_B}L^{1/\nu}$
respectively. Both figures indicate data collapse onto a single curve for the indicated values of $T^{\text{cep}}$, 
$\mu_B^{\text{cep}}$ and $\nu$, albeit with much less statistical significance for $\kappa\sigma^2$ (cf. Fig.~\ref{Fig2}), 
especially for $T$ (${\mu_B}$) values smaller (larger) than $T^{\text{cep}}$ ($\mu_B^{\text{cep}}$). 
The resulting scaling functions for  $S\sigma$ and $\kappa\sigma^2$ also indicate characteristic shape differences
akin to those for the 3D Ising model.

The scaling functions in Figs.~\ref{Fig1} and \ref{Fig2} point to the efficacy of Finite-Size Scaling as a tool to 
locate and characterize the CEP. They indicate that the non-Gaussian fluctuations for net-protons, 
quantified by $S\sigma$ and $\kappa\sigma^2$ respectively, are consistent with the critical fluctuations expected for 
reaction trajectories close to a second order phase transition at a CEP located at $T^{\text{cep}} \sim 165$~MeV 
and $\mu_B^{\text{cep}} \sim 95$~MeV, belonging  to the 3-D Ising Model universality class~\cite{Lacey:2014wqa,Lacey:2015yxg}.

It is often argued that theoretical and experimental estimates limits the location of the CEP to $\mu_B^{\text{cep}}$ values larger 
than 200 MeV \cite{Luo:2015doi,Karsch:2015nqx}.
Consequently, it is instructive to perform scaling tests with $\mu_B^{\text{cep}}$ and  $T^{\text{cep}}$ values 
different from the ones used for the scaling tests shown in Figs.~\ref{Fig1} and \ref{Fig2}, {\em i.e.}, different from the values 
reported in Ref.~\cite{Lacey:2014wqa}.  The results from one such test are shown in Fig.~\ref{Fig3} for  
the {\em test} values $\acute{T}^{\text{cep}} \sim 160$~MeV and $\acute{\mu_B}^{\text{cep}} \sim 205$~MeV. These values
correspond to chemical freeze-out  at $\sqrt{s_{NN}} \sim 20$ GeV, where minima/maxima 
in the non-monotonic patterns for several observables have been 
reported~\cite{Lacey:2014rxa,Adare:2014qvs,Luo:2015doi,Adamczyk:2014ipa,Adamczyk:2016exq}.  

In contrast to Figs.~\ref{Fig1} and  \ref{Fig2}, Fig.~\ref{Fig3} clearly shows that Finite-Size Scaling is broken,
indicating that the {\em test}  values $\acute{T}^{\text{cep}}$ and $\acute{\mu_B}^{\text{cep}}$  do not constitute a good 
estimate for the location of the CEP.  Similar results were obtained  for 
other $\acute{T}^{\text{cep}}$ and $\acute{\mu_B}^{\text{cep}}$ values, corresponding to even 
smaller values of  $\sqrt{s_{NN}}$, {\em i.e.}, smaller values of $\acute{T}^{\text{cep}}$ and  larger 
value of $\acute{\mu_B}^{\text{cep}}$.

%
%
In summary, we have performed validation tests for effective Finite-Size Scaling of 
recent STAR measurements of  the excitation function for non-Gaussian fluctuations, characterized by the 
moment products $S\sigma $  and  $\kappa \sigma^2 $ of  the event-by-event net-proton
multiplicity distributions. The  scaling  tests, which provide a potent tool for 
locating and characterizing the CEP, validate characteristic FSS patterns consistent 
with a second order deconfinement phase transition at the critical end point. They also indicate consistency 
with the earlier estimate~\cite{Lacey:2014wqa,Lacey:2015yxg} of $T^{\text{cep}} \sim 165$~MeV and 
$\mu_B^{\text{cep}} \sim 95$~MeV for the location of the CEP,  and  the critical exponents  used 
to assign its 3D Ising Model universality class. Further detailed studies at RHIC, with the slated STAR detector 
upgrades and RHIC beam intensity enhancements -- especially for the low energy beams, are  
crucial to firm-up the  precise location of the CEP,  and the values of the critical exponents required 
for its precise characterization.

\section*{Acknowledgments}
The authors thank Jiangyong Jia, and  Zhangbu Xu for valuable discussions. 
This research is supported by the US DOE under contract DE-FG02-87ER40331.A008.

\bibliography{ref_fss_fluc_ms}   

\begin{thebibliography}{40}
\bibitem[{exp()}]{exponents}
\bibinfo{note}{The critical exponents satisfy several equalities. Thus, only
  two of the four exponents are independent.}
\end{thebibliography}

\end{document}